# Study of All–Fiber Asymmetric Interleaver Based on Mach-Zehnder Interferometer

Lu Huaiwei[a], Wei Yun[a], Zhang Baoge[b], Wu Kaijun[a], Luo Guanwei[a]
([a.]School of Mathematics, Physics and software Engineering, Lanzhou Jiaotong University, Lanzhou 730070, China，
[b.]School of Automation and Electrical Engineering, Lanzhou Jiaotong University, Lanzhou 730070, China)

**Abstract:**
In order to improve the transmission efficiency of optical-fiber communication system with 10Gb/s+40Gb/s, an all-fiber interleaver with unequal passband is proposed and discussed, which is based on a two-stage cascaded Mach-Zehnder interferometer (MZI). The optimum value of structural parameters, such as splitting ratios of the couplers and the physical length differences of the interferometer arms, were chosen. One set of optimized data is validated in the experimental result.The experimental results and the theoretic analysis indicate that an all-fiber optical interleaver with -3dB passband width in odd channels and even channels could be obtained, which having more than 60GHz passband and 30GHz passband, for transmission speed of 40Gb/s and 10Gb/s, respectively.By assigning different portions of spectrum to the 10Gb/s and the 40 Gb/s channels, the bandwidth efficiency requirement of the 40 Gb/s channel is relieved, and therefore longer transmission distance can be achieved.Comparing with the conventional asymmetric interleaver, the most obvious benefit of the proposed interleaver is all-fiber structure.

## 1. Introduction

To accommodate the ever increasing network traffic, network providers are doubling the channel number to 50GHz spaced system and adding high bit rate channels such as OC-768/STM256 at 40 Gb/s. At the same time, it is desirable to maintain the legacy 10 Gb/s systems to keep the network upgrade capital expense low. Therefore dense wavelength division multiplexing (DWDM) system with spectrally alternating 10G and 40G channels is a suitable scheme. And optical interleavers are used for inserting and extracting the 40 Gb/s channels. However, the 50 GHz bandwidth provided by conventional symmetric interleaver is not sufficient for transmitting 40 Gb/s signal at long distance, especially for modulation schemes that are not spectrally efficient. For optical fiber DWDM transmission systems with 10Gb/s+40Gb/s, the researching and designing of asymmetric interleaver is very essential[1].

As it serves as a component in optical fiber DWDM transmission systems, an optical interleaver is highly desired to have low dispersion, low insertion loss and low crosstalk. But more importantly, it should be well compatibility with optical fiber DWDM transmission systems. With the rapid development of the fused biconical taper (FBT) technology, fused cascade MZI is highly concerned in the application to the symmetrical interleaver,but there are seldom reports on the application and study in the asymmetric interleaver [2~4].

In this paper, an all-fiber asymmetric optical interleaver based on a two-stage cascaded MZI is proposed.The results of numerical simulation and experiments show that the proposed interleaver with -3dB passband in odd channels and even channels could be obtained, which having more than 60GHz passband and 30GHz passband, for transmission speed of 40Gb/s and 10Gb/s, respectively. And Its asymmetric interleaving feature provides an ideal solution for this unbalanced DWDM system.Comparing with the conventional asymmetric interleaver, the biggest advantage of the proposed interleaver is all-fiber structure.

## 2. Device structure and design

Fig.1 shows the configuration of the all-fiber asymmetric optical interleaver based on a two-stage cascaded MZI. The device consists of three 2×2 single mode fiber couplers $DC_j$ ($j$=1,2,3) which are connected with fiber arms $l_1$, $l_2$, $l_3$ and $l_4$. These arms $l_1$, $l_2$, $l_3$ and $l_4$ are the same single mode fiber.For the convenience of derivation, $E^{ini}$ ($i$ = 1, 2 ) and $E^{outi}$, respectively,are expressed as the input field and output field.

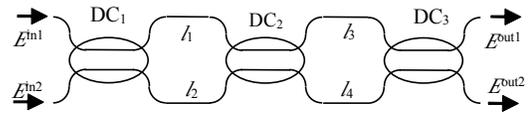

Fig.1 Schematic diagram of the all-fiber asymmetric interleaver

Because the light transmission distance of the optical fiber in the device is quite short, the transmission loss may be neglected. Assuming that the input optical field $E^{in}$($E^{in1}=E^{in}$ and $E^{in2}=0$) is inputted from the upper port of $DC_1$. And $E^{in}$ is a linearly polarized light. After $E^{in}$ travelled through $DC_1$, it is split into two signals.These signals are traveling direction through $l_1$, $l_2$, $l_3$ and $l_4$, respectively, and will interfere in the couplers $DC_2$ and $DC_3$.The normalized output intensities $T_1(\theta)$ ($=|E^{out1}|^2/|E^{in}|^2$) and $T_2(\theta)$ ($=|E^{out2}|^2/|E^{in}|^2$) can be expressed as follows

$$\begin{cases} T_1(\theta) = a_0 + a_1\cos(\theta) + a_2\cos(2\theta) \\ T_2(\theta) = b_0 + b_1\cos(\theta) + b_2\cos(2\theta) \end{cases} \quad (1)$$

Here

$a_1 = 2^{-1}\sin(2K_2)\sin(2K_1+2K_3) = -b_1$

$a_2 = 2^{-1}\sin(2K_1)\cos^2(K_2)\sin(2K_3) = -b_2$

$a_0 = [1-\sin(2K_1)\sin^2(K_2)\sin(2K_3)-\cos(2K_1)\cos(2K_2)\cos(2K_3)]/2$

$b_0 = [1+\sin(2K_1)\sin^2(K_2)\sin(2K_3)+\cos(2K_1)\cos(2K_2)\cos(2K_3)]/2$

$\theta=\beta(l_2-l_1)=\beta(l_4-l_3)=2\pi n_{\text{eff}}\Delta l/\lambda$.$\beta$ is the propagation constant

in the fiber. $\lambda$ is the wavelength and $n_{\text{eff}}$ is the effective index of the fiber. $K_j$ are the coupling-coefficient-angles of $DC_j$, which is equal to the multiplier of the coupling coefficient and the equivalent coupling length of the coupling region of a single mode fiber coupler.

Generally speaking, all optical interleavers are made of optical interferometer. The interference creates an optical output which is a periodic function of the frequency. For the structure of all-fiber asymmetric interleaver based on a cascaded MZI with three directional couplers, the function of input coupler $DC_1$ is distribution of the input light, the one of the other two couplers $DC_2$ and $DC_3$ is interference of the light beam respectively. The period and output waveforms of an optical interleaver are determined by the path length difference in the arms of the interferometer, as well as the splitting ratios of the fiber couplers. High adjacent channel isolation is the basic requirements for an optical interleaver.

According to Fourier series theory, output waveform can be constituted with a series of sines or cosines corresponding to Fourier relations. It can be seen from Eq.1 that $T_i(\theta)$ are all periodical with respect to $\theta$ or the incident wavelength $\lambda$, and are characterized by a sum of simple sinusoidal components corresponding to Fourier relations. The coefficients $a_i$ and $b_i(i=0\sim2)$ are the function associated with $K_j$, which determine the output waveforms of $T_i(\theta)$. Therefore the interleaver design can be described as to obtain appropriate $K_j$. By straightforward calculations, the first derivatives of $T_i(\theta)$ follows that

$$\begin{cases} \dfrac{dT_1(\theta)}{d\theta} = T_1'(\theta) = -\sin(\theta)[a_1 + 4a_2\cos(\theta)] \\ \dfrac{dT_2(\theta)}{d\theta} = T_2'(\theta) = -\sin(\theta)[b_1 + 4b_2\cos(\theta)] \end{cases} \quad (2)$$

Forcing $T_i'(\theta)$ equal to zero, it is derived that $T_i(\theta)$ has stationary points in $\theta_1 = 0 \pm n\pi$ ($n$ is a positive integer) and $\theta_{2,3} = \pm\cos^{-1}[-a_1/(4a_2)]$. The second derivatives of $T_i(\theta)$ at the extreme points of $\theta_1$ and $\theta_{2,3}$ are as follows:

$$\begin{cases} \left.\dfrac{d^2T_1(\theta)}{d\theta^2}\right|_{\theta=0} = T_1''(0) = -[a_1+4a_2] \\ \left.\dfrac{d^2T_1(\theta)}{d\theta^2}\right|_{\theta=\theta_{2,3}} = T_1''(\theta_{2,3}) = \dfrac{-a_1^2}{4a_2} + 4a_2 \\ \left.\dfrac{d^2T_2(\theta)}{d\theta^2}\right|_{\theta=0} = T_2''(0) = [a_1+4a_2] \\ \left.\dfrac{d^2T_2(\theta)}{d\theta^2}\right|_{\theta=\theta_{2,3}} = T_2''(\theta_{2,3}) = \dfrac{a_1^2}{4a_2} - 4a_2 \end{cases} \quad (3)$$

It can be seen from Eq.3 that both $T_1''(0)$ and $T_2''(0)$ are just the opposite, and it is the same with $T_1''(\theta_{2,3})$ and $T_2''(\theta_{2,3})$. This means that when $T_1(0)$ and $T_1(\theta_{2,3})$ is the maximum (or the minimum), the corresponding extremum value for $T_2(0)$ and $T_2(\theta_{2,3})$ is the minimum (or the maximum). Substituting $\theta=0$ and $\theta=\pi$ into (1), they are given by

$$\begin{cases} T_1(0) = \dfrac{1-\cos[2(K_1+K_2+K_3)]}{2} \\ T_2(0) = \dfrac{1+\cos[2(K_1+K_2+K_3)]}{2} \\ T_1(\pi) = \dfrac{1-\cos[2(K_1-K_2+K_3)]}{2} \\ T_2(\pi) = \dfrac{1+\cos[2(K_1-K_2+K_3)]}{2} \end{cases} \quad (4)$$

It is showed that $T_1(0)=1$, $T_2(0)=0$ at $K_1+K_2+K_3=\pi/2$ and $T_1(\pi)=0$, $T_2(\pi)=1$ at $K_1-K_2+K_3=0$. Obviously, in order to meet $K_1+K_2+K_3=\pi/2$ and $K_1-K_2+K_3=0$ simultaneously, $K_1,K_2$ and $K_3$ should be $K_2=\pi/4$ and $K_1+K_3=\pi/4$.

Substituting $K_1=K$, $K_2=\pi/4$ and $K_3=\pi/4-K$ (or $K_3=K, K_1=\pi/4-K$) into Eq.(1), Eq.(1) is rewritten as

$$\begin{cases} T_1(\theta) = \dfrac{1}{2} + \dfrac{\cos(\theta)}{2} + \dfrac{\sin(4K)(\cos(2\theta)-1)}{8} \\ T_2(\theta) = \dfrac{1}{2} - \dfrac{\cos(\theta)}{2} - \dfrac{\sin(4K)(\cos(2\theta)-1)}{8} \end{cases} \quad (5)$$

For optical-fiber communications purposes, the channel crosstalk is one of the most important factors which block the wide use of an interleaver. The interleaver design problem is reduced to finding those splitting ratios to minimize the channel crosstalk. It is equivalent to maximizing the channel isolation given by $\text{Max}|T_1(0)-T_1(\pi/2)|$ or $\text{Max}|T_2(0)-T_2(\pi/2)|$ [5], that is,

$$\begin{cases} \text{Max}|T_1(0)-T_1(\pi/2)| = \dfrac{1}{2}[1+\dfrac{\sin(4K)}{2}] \\ \text{Max}|T_2(0)-T_2(\pi/2)| = \dfrac{1}{2}[1+\dfrac{\sin(4K)}{2}] \end{cases} \quad (6)$$

Apparently, Eq.(6) will reach the maximum when $K=\pi/4$. So far we have obtained optimization design parameters $K_1=K_3=\pi/8$ and $K_2=\pi/4$ for the proposed interleaver in accordance with the maximum channel isolation.

To verify the theoretical analysis, simulation studies are carried out for the design of the all-fiber asymmetric interleaver which is capable of separating two sets of interleaving 50 GHz signals. In the design, the differential delay $\Delta l$ is tuned for 50GHz.

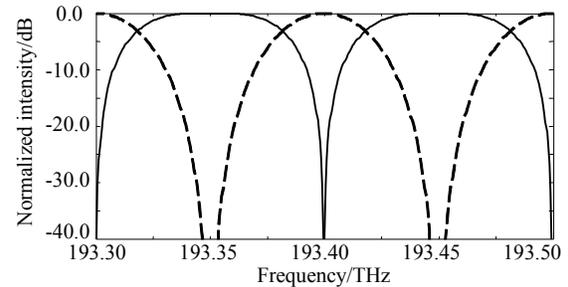

Fig.2 The spectra of the all-fiber asymmetric interleaver with $K_1=K_3=\pi/8$ and $K_2=\pi/4$.

The following simulation results are calculated utilizing Eq.(1) with the different parameters $K_j$. The solid line represents $T_1(\theta)$, the dotted line $T_2(\theta)$, and $\Delta f$ the -3dB passband width respectively. The simulation results with optimization design parameters $K_1=K_3 = \pi/8$ and $K_2=\pi/4$, is shown in Fig.2. In Fig.2, $\Delta f$ of $T_1(\theta)$ and $T_2(\theta)$ are 62.56GHz and 35.66GHz respectively. Their channel isolations on two output port are more than 40dB, too. Obviously, $\Delta f$ and the corresponding transmission rate $v$ are accorded with the most appropriate relationship $\Delta f \geq 1.5v$. Therefore, the proposed interleaver can be applied to 10GHz+40GHz hybrid system. And by assigning different portions of spectrum to the 10Gb/s and the 40 Gb/s channels, the bandwidth efficiency requirement of the 40 Gb/s channel is relieved, and therefore longer transmission distance can be achieved.

The asymmetric interleavers in [5,6] are all-fiber structures, but the interleaver in [5] requires two segments of high birefringence fiber, length ratio 1:2, in a sagnac loop mirror. And the other one in [6] need to add 33 couplers in series with various coupling ratio. The channel isolations of both asymmetric interleavers in[5,6] are slightly higher than 30dB. Comparing with these asymmetrical interleavers in [5,6], the proposed interleaver may not only realize the asymmetrical bandwidth output but also improve the channel isolation.

## 3.Tolerance analysis

The channel crosstalk and isolation are the most sensitive to the splitting ratios of couplers, and the splitting ratios tolerance corresponds to the fabrication process tolerance.In the fabrication process, the parameters of a fused cascade MZI may deviate from the designed values due to fabricating process error. One process error is a deviation from the expected splitting ratios of couplers $DC_j$. These deviation are equivalent to the deviation from of $K_j$. So, the relation between the channel isolation and $K_j$ should be discussed.

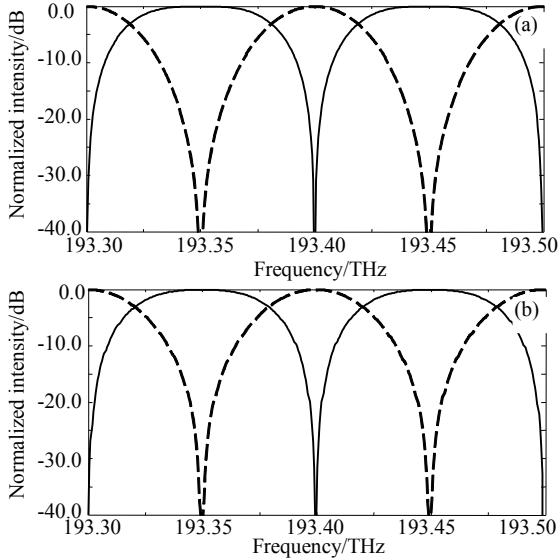

Fig.3 The spectra of the interleaver with $K_2=\pi/4$ and (a) $K_1=\pi/12$, $K_3=\pi/4-K_1$ (b) $K_1 =\pi/5$, $K_3=\pi/4-K_1$.

In Fig.3, let's assume $K_2=\pi/4$ unchanged. Thus, the analysis is focused on $K_1$ and $K_3$. The simulation results of the simulation results of Eq.(1) with the parameters $K_1= \pi/12$, $K_2=\pi/4$ and $K_3=\pi/4-K_1$ is shown in Fig.3(a), $\Delta f$ of $T_1(\theta)$ and $T_2(\theta)$ are 60.86GHz and 36.16GHz respectively. And that of the Eq.(1) with the parameters $K_1=\pi/5$, $K_2=\pi/4$ and $K_3=\pi/4-K_1$ is shown in Fig.3(b), $\Delta f$ of $T_1(\theta)$ and $T_2(\theta)$ are 62.88GHz and 34.86GHz respectively. In Fig.3(a) and (b),the channel isolations are over 30dB. Comparing with calculated results of Fig.2, $K_1<\pi/8$(and $K_2>\pi/8$) is corresponding to the decrease of $\Delta f$ of $T_1(\theta)$ and the increase of $\Delta f$ of $T_2(\theta)$. On the contrary,$K_1>\pi/8$(and $K_2<\pi/8$) is corresponding to the decrease of $\Delta f$ of $T_2(\theta)$ and the increase of $\Delta f$ of $T_1(\theta)$.

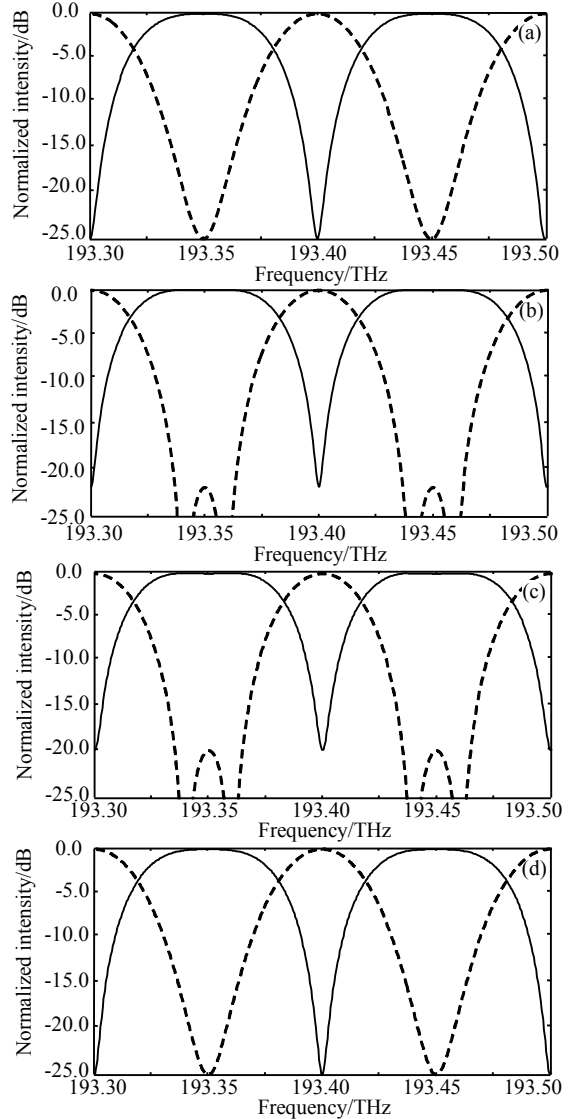

Fig.4 The spectra of the interleaver with (a)$K_1=K_3=\pi/8$, $K_2= 2\pi/7$, (b)$K_1=K_3=\pi/8$, $K_2=2\pi/9$ (c)$K_2=\pi/4$, $K_1=K_3= \pi/7$,(d)$K_2=\pi/4$, $K_1= K_3 =\pi/9$

Keeping $K_1=K_3=\pi/8$ unchanged, the simulation results of Eq.(1) with the parameters $K_2=2\pi/7$ and $K_2=2\pi/9$ are shown in Fig.4(a) and (b) respectively.In Fig.4(a), $\Delta f$ of $T_1(\theta)$ is 64.56GHz and $\Delta f$ of $T_2(\theta)$ is 33.86GHz. And in Fig.4(b), $\Delta f$ of $T_1(\theta)$ is 60.55GHz and $\Delta f$ of $T_2(\theta)$ is 37.86GHz.Comparing with calculated results of Fig.2, the $\Delta f$ of $T_1(\theta)$ is widened and $\Delta f$ of $T_2(\theta)$ is narrowed when $K_2 < \pi/4$, and the $\Delta f$ of $T_1(\theta)$ is narrowed and $\Delta f$ of $T_2(\theta)$ is widened when $K_2>\pi/4$. Thus it could be concluded that the channel isolations are also very noticeable, whatever the splitting ratio of coupler $DC_2$ is greater than or less than 0.5:0.5

By useing the similar computational method, keeping $K_2=\pi/4$ unchanged, Fig.4(c) and (d) are the simulation results of Eq.(1) with the parameters $K_1=K_3=\pi/7$ and $K_1=K_3=\pi/9$,which no longer no longer satisfy $K_1+K_3=\pi/4$, respectively. In Fig.4(c), $\Delta f$ of $T_1(\theta)$ is 62.56GHz and $T_2(\theta)$ is 33.36GHz. And in Fig.4(d), $\Delta f$ of $T_1(\theta)$ is 61.06GHz and $T_2(\theta)$ is 37.36GHz.

Apparently, the simulation results of Fig.4 show that the channel isolations are also very noticeable, though $\Delta f$ and $v$ are satisfied with $\Delta f \geq 1.5v$.From the above we can come to the conclusion that the main cause for declining the channel isolation are mainly due to $K_2 \neq \pi/4$ and $K_1+K_3 \neq \pi/4$. Therefore, the key technologies for an all-fiber asymmetric interleaver based on a cascaded MZI is to accurately control $K_2$. On the basis of coupler $DC_2$ being a 3dB coupler, $K_1$ and $K_3$ should be as close as possible to meet $K_1+K_3=\pi/4$. In other words, if the splitting ratio of coupler $DC_1$(or $DC_3$ )is greater than(or less than) 15:85, then the that of coupler $DC_3$(or $DC_1$) should be preferably less than (or greater than) 15:85.

## 4. Experiment procedure

The proposed approach has been used to design an all－fiber asymmetric interleaver based on a cascaded MZI. The interleaver is fabricated by fusing two pieces of single-mode fibers.The fabrication method proposed in [7,8] is employed to fabricate the interleaver without any splicing. In the course of fabricating the interleaver, the fused biconical taper technology is adopted.

For the above the discussion,the accuracy of the splitting ratio of coupler $DC_2$ is more important to power transmission than that of couplers $DC_1$ and $DC_3$.And the accuracy of splitting ratio of the coupler $DC_2$ is crucial to fabricate all-fiber asymmetric interleaver based on a cascaded MZI. It is highly necessary to keep the power splitting ratio of coupler $DC_2$ close to 50:50 as possible. In order to control the precision of the splitting ratio of coupler $DC_2$, the first step is to fabricate the coupler $DC_2$. A broadband super-fluorescent EDF source, with a nearly flat spectral response in the range 1548-1558nm, is used as a monitoring power source.In this way,the coupler $DC_2$ can be easily made under the power monitoring to ensure that the power coupling ratio for two output ports is 50:50.However, there is some trouble in fabricating the fiber couplers $DC_1$ and $DC_3$. The power splitting ratio of 15:85 can be approximately realized only by strictly through the experience of the technician. $DC_i$ are made by the same single-mode fiber, and the splitting ratios are controlled by a microcomputer and an oxygen-hydrogen burner by electronic mass flow controllers.

The fibers linking the couplers are initially obtained with roughly the estimated lengths to meet the differential delay requirement. After the fabrication of the couplers, all two fibers on one arm are then stretched to achieve the precise length difference by monitoring the interference spectrums by using an optical spectrum analyzer(OSA).Once all the desired differential delays are obtained, the phase shifts can be achieved by using heaters on the sections of the differential arms until the desired interleaving pattern is observed [8,9].By this technique, the fiber arms of MZI was adjusted until the desired interleaving pattern is observed.

In process of the fabricating all-fiber asymmetric interleaver based on a cascaded MZI, it is difficult to avoid the stochastic change of polarization of a transmitted signal due to many factors such as the fiber bending, physical defect of fiber itself, etc. A method using polarization squeezer to control the drift on state of polarization in the fiber interferometer arms.Another method is to shorten the fiber length as much as possible in order to reduce the polarization's drift caused by the lengths.In this experiment, the length of fiber interferometer arm is controlled under two meters.Moreover,during fabricating procedure care must be taken that two sections of single mode fiber are not under any pressure tension or twisting in order to maintaining polarization stabilization.

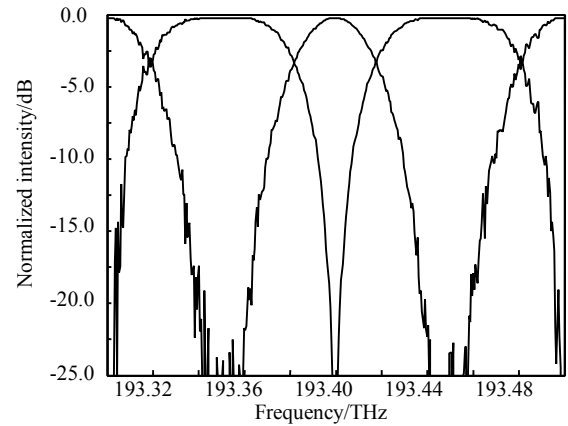

Fig.5 Measured Experimental spectrum of the all-fiber unequal-passband interleaver

To evaluate the performance of the obtained all-fiber asymmetric interleaver, the output spectrums in the response to the broad-band light sent at the input port, are observed by using an OSA. A tunable laser with wavelength between 1540nm-1560nm is used as the input light.Through the polarizer,the input light is changed into the linearly polarized light, and it is inputted from the upper port of $DC_1$.The experiments were done most carefully again and again. The optical performance of the obtained all-fiber asymmetric interleaver is measured by Agilent 81910A Photonics All-parameter Analyzer. The obtained experimental data are shown in Fig.5. It is clear that the designed interleaver, in 50-GHz channel spacing, with asymmetric $\Delta f$ in odd channels and even channels could be obtained, which having 60.16GHz and 31.32GHz passband width respectively. The average channel isolation is better than 25 dB. The average insertion loss of the proposed component is estimated to be about 0.16dB. The experimental result shows that the proposed configuration and design approach are effective for designing all-fiber asymmetric optical interleaver.

When the experiment is completed, the splitting ratios of couplers $DC_i$ are detected by cutting the fiber interferometer arms. The measure values of the splitting ratios of $DC_1$, $DC_2$ and $DC_3$ are 20:80, 50:50 and 10:90 respectively. It shows that the main cause for channel isolation decline in Fig.5 are mainly due to the splitting ratio deviations of couplers $DC_2$ and $DC_3$.

## 5. Conclusion

An all-fiber asymmetric optical interleaver based on a two-stage cascaded MZI is proposed and discussed. Influences on its transmission characteristics by such factors as splitting ratio of the coupler of the directional fiber couplers and the physical length differences of the interferometer arms are numerically analyzed in detail. The preferences condition of the interleaver which meets the maximal isolation and the characteristic of the output spectrum is obtained. One set of optimized data is validated in the experimental result. The results of numerical simulation and experiments show that the obtained all-fiber optical asymmetric interleaver with the different -3dB passband width in odd channels and even channels could be obtained, which having more than 60GHz and 30GHz passband width respectively. For 10Gb/s+40Gb/s, the -3dB passband width offered by the obtained all-fiber asymmetric optical interleaver leads to superior overall performance across all channels comparing to symmetric optical interleaver, regardless of the 40Gb/s channel modulation scheme. Therefore, the obtained all-fiber asymmetric optical interleaver based on a two-stage cascaded MZI has potential applications foreground in optical fiber DWDM transmission systems with 10Gb/s+ 40Gb/s10.

## Acknowledgement

This study is supported by National Natural Science Foundation project of China(No.10972095) and Gansu of China(No. 0803RJ ZA027).

## References

[1].L.Ceuppens,T.Schmidt, A. Liang, and K.-P. Ho.Proc. NFOEC,(2003) 546
[2].S.W.Kok, Y.Zhang, C.Y.Wen, Y.C.Soh, Opt. Commun.226(2003)241.
[3].Q.Wang, S.L.He. J. Lightw. Technol, 23(2005) 1284
[4].H.W.Lu,B.G.Zhang,K.J.Wu,Y.Wei,G.W.Luo.Acta Optica Sinica,30 ( 2010) 2406
[5]. R.F.Zhang, S.H.Wang,C.F.Ge. J. Tianjin University, , 39(2006) 365.
[6]. T.Zhang, K.Chen, S.A.Zhao. J.Optoelectronics·Lase,16(2005) 436.
[7].H.W.Lu,B.G.Zhang,M.Z.Li,W.G.Luo.IEEE Photon. Technol. Lett. 18 (2006)1469.
[8].Q.J. Wang, Y. Zhang, Y.C. Soh, IEEE. Photon. Technol. Lett. 16(2004) 168.
[9].H.W.Lu, Y.E.Zhang, G.W. Luo.Opt. Commun.276 (2007) 116